\documentclass{article}
\usepackage{ijcai24}

\usepackage{times}
\usepackage{soul}
\usepackage{url}
\usepackage[hidelinks]{hyperref}
\usepackage[utf8]{inputenc}
\usepackage[small]{caption}
\usepackage{graphicx}
\usepackage{amsmath}
\usepackage{amsthm}
\usepackage{booktabs}
\usepackage{algorithm}
\usepackage{algorithmic}
\usepackage[switch]{lineno}

\usepackage{amsfonts}
\usepackage{tabularray}
\usepackage{amsmath}
\usepackage{subfigure}
\usepackage{color}

\title{CGAP: Urban Region Representation Learning with Coarsened Graph Attention Pooling}

\author{
Zhuo Xu$^1$
\and
Xiao Zhou$^{2}$  \thanks{Xiao Zhou is the corresponding author.}\\
\affiliations
$^1$School of Artificial Intelligence, Beijing Normal University\\
$^2$Gaoling School of Artificial Intelligence, Renmin University of China\\
\emails
xiaozhou@ruc.edu.cn
}

\begin{document}

\maketitle

\begin{abstract}
    The explosion of massive urban data recently has provided us with a valuable opportunity to gain deeper insights into urban regions and the daily lives of residents. Urban region representation learning emerges as a crucial realm for fulfilling this task. Among deep learning approaches, graph neural networks (GNNs) have shown promise, given that city elements can be naturally represented as nodes with various connections between them as edges. However, many existing GNN approaches encounter challenges such as over-smoothing and limitations in capturing information from nodes in other regions, resulting in the loss of crucial urban information and a decline in region representation performance. To address these challenges, we leverage urban graph structure information and introduce a hierarchical graph pooling process called Coarsened Graph Attention Pooling (\textbf{CGAP}). CGAP features local attention units to create coarsened intermediate graphs and global features. Additionally, by incorporating urban region graphs and global features into a global attention layer, we harness relational information to enhance representation effectiveness. Furthermore, CGAP integrates region attributes such as Points of Interest (POIs) and inter-regional contexts like human mobility, enabling the exploitation of multi-modal urban data for more comprehensive representation learning. Experiments on three downstream tasks related to the UN Sustainable Development Goals validate the effectiveness of region representations learned by our approach. Experimental results and analyses demonstrate that CGAP excels in various socioeconomic prediction tasks compared to competitive baselines.
\end{abstract}

\section{Introduction}


In recent years, there has been a rapid growth in urban data collection, driven by urbanization and the proliferation of mobile devices~\cite{urban_data_explosive}. Cities are composed of diverse regions where residents live, work, and engage in various activities. Optimizing the representation of urban regions can unveil internal properties and correlations within cities. Effective dense representations of urban regions can facilitate the prediction of various urban computing tasks, including criminal prediction~\cite{criminal_prediction}, traffic prediction~\cite{traffic_prediction}, and can inform policy-making~\cite{zhou2017cultural,zhou2023phase}, urban planning~\cite{zhou2018discovering,zhou2019topic,wang2018predicting}, and sustainable development~\cite{zhang2024causally}. Despite the emergence of large amounts of urban data, such as Points of Interest (POIs) and population mobility, and the growing demand for region representation learning, effectively learning representations of urban regions from such vast datasets remains a significant challenge.


Previous methods have endeavored to leverage the spatial characteristics of regions and represent urban areas as structured graphical data~\cite{graph_structure0,graph_structure1,graph_structure2}. Utilizing this graphical structure, early studies have applied graph representation learning techniques such as Graph Neural Networks (GNNs) to acquire embeddings for regions~\cite{gnn_0,gnn_1}. For instance, \citeauthor{HREP} \shortcite{HREP} employed relation-aware Graph Convolutional Networks (GCNs) to learn embeddings for both regions and relation types. Additionally, many urban region attributes exhibit temporal and spatial characteristics. GNNs equipped with spatio-temporal feature recognition have been employed for region embedding~\cite{temporalGNN_0,temporalGNN_1,temporalGNN_2}. For instance, \citeauthor{temporalGNN_3} \shortcite{temporalGNN_3} devised a spatio-temporal GNN that integrates GNNs with multi-step dependency relations to support applications in spatio-temporal prediction within urban environments.


While graph representation learning methods have shown promise in urban computing tasks, they still grapple with major limitations inherent in GNN architectures: (1) information propagation limited to edges, resulting in flat networks~\cite{flat}; (2) region nodes capable of learning their feature embeddings but failing to capture information from other regions in the graph structure~\cite{HREP}. Addressing these dual challenges within vanilla GNN frameworks remains difficult for region representation learning tasks. Some approaches attempt to enhance inter- and intra-region interaction modeling within GNN architectures but struggle with leveraging the full multi-modal information within urban environments~\cite{urban2vec}. Despite efforts to supplement regional features with diverse data, simply applying operations on the data cannot compensate for structural deficiencies in the model. Furthermore, while existing studies have demonstrated the utility of urban global features in downstream tasks~\cite{global_fea0,MGFN}, current graph representation learning methods fail to effectively integrate such global information~\cite{MGFN}. For instance, \citeauthor{MGFN}~\shortcite{MGFN} proposed a joint learning approach that leverages inter- and intra-pattern information to enhance region representation. However, the integration of global features remains a challenge for existing graph representation learning methods.

To tackle the challenges of graph representation learning in urban region embedding, our focus lies in achieving effective information propagation among local region nodes within GNN architecture. Building on insights from existing studies demonstrating the utility of urban global features in downstream tasks~\cite{global_fea0,MGFN}, we propose Graph Neural Networks with Coarsened Graph Attention Pooling (\textbf{CGAP}). CGAP captures both local region features and urban global features through multi-level pooling. Specifically, it integrates global and regional features to extract graphical node embeddings, aiming to capture the interaction characteristics of the entire urban area through multi-layer pooling and incorporate them with regional features via a global attention mechanism.
Our proposed framework comprises three sequential modules: (1) a GNN module for obtaining the original embedding of the urban region graph; (2) CGAP, which includes multi-layer pooling and local attention units to generate coarsened graphs and global features from the original graph; (3) a global attention layer for integrating global features and original graph embeddings. Additionally, we define learning objectives related to urban computing downstream tasks to optimize learning at the end of the framework. Furthermore, urban regions exhibit diverse attributes and corresponding data, necessitating the consideration of multiple factors during training. Thus, our model, leveraging CGAP, integrates these various attributes into its learning objectives.

The key contributions of our work can be summarized as follows:

\begin{itemize}

\item We formalize urban region embedding as a graph representation learning task, integrating diverse urban data sources to capture complex interwoven region correlations and features for more effective urban region representation learning.


\item We propose CGAP, a novel method that enhances GNNs for region representation learning by capturing both intra- and inter-region interactions through multi-layer pooling, generating urban global features and improving representation effectiveness.


\item Through comprehensive experiments, we validate the effectiveness of our method across various real-world urban datasets and downstream prediction scenarios, outperforming competitive baseline models and demonstrating its capability in capturing intricate urban dynamics and enhancing predictive performance.

\end{itemize}


The remainder of our paper is organized as follows. Firstly, we provide problem-related definitions and formulate the problem. Next, we delve into the details of our framework, explaining the CGAP mechanism and demonstrating its effectiveness through experiments on real-world datasets. Following this, we compare our model with state-of-the-art methods. Finally, we review related works and conclude the paper.

\section{Problem Definition}

\label{problem_definition}This research aims to learn representations of urban regions by leveraging the urban graph structure and its properties.  The objective is to train region embeddings that are effectively encoded from the region information, thereby benefiting multiple downstream  tasks in the urban computing.

\subsection{Urban Region Graph}
An urban is composed of regions that are connected. Each region has its own internal natural geographic or city attributes. With the exchange of residents between regions, the relationship between regions is formed. Therefore, the city is represented as $\mathcal{G} = (\mathcal{V},\mathcal{E})$ where $\mathcal{V}$ is the vector set to represent regions and $\mathcal{E}$ is the edge set. The geographic neighbor information of $\mathcal{G}$ is represented as $N = \{n_{ij}\} (n_{ij} \in \{0,1\}, \forall{i, j}\in [1,\|\mathcal{V}\|])$ which is the adjacency matrix.

\subsection{Region Attributes}
The region attributes are the geographic and social features of urban regions. Especially, Point-of-Interests (\textbf{POI}s)  are the most important regional attributes that we focus on. POI such as a shop, station, or hospital is any meaningful point (except geographical meaning) on the map. We define region attributes as $P = \{p_1, p_2, ...,p_{\|\mathcal{V}\|} \}$ where $p_{i}$ is the number of POI in region $i$ and $\|\mathcal{V}\|$ is the number of regions.

The interaction among regions relies on the movement of residents. Given an urban region graph $\mathcal{G}$, human mobility is defined as a directed and weighted graph, and its adjacency matrix is represented as $M = \{m_{ij}\}$ where $m_{ij}$ represents the number of residents moving from region $v_{i}$ to $v_{j} ( \forall{v_i,v_j}\in \mathcal{V})$. Generally, we utilize a combination of POIs and mobility data and define it as multi-view data, which contains abundant city information.

\subsection{Urban Graph Representation Learning}
Given an urban region graph $\mathcal{G}$,  neighbor information $N$, region attribute $P$, and region relation $M$, our objective is to learn the embedding of urban region graph $E = \{e_1, e_2, ..., e_{\|\mathcal{V}\|}\}$ where $e_{i} \in \mathbb{R}^{d}$ is the embedding result of region $i (\forall{i} \in [1, \|\mathcal{V}\|])$, $d$ is embedding dimension. This learning task can be represented as follows:
\begin{equation}
    \Gamma : \chi \rightarrow  E \in \mathbb{R}^{\|\mathcal{V}\| \times d},
\end{equation}
where $\chi = \langle \mathcal{G}, N, P, M \rangle$, each row in $E$ is $e_i$ and $\Gamma (\cdot)$ is the function that encodes multiple information related to city into the latent space for region representation. The urban computing downstream tasks utilize the embedding results as input to analyze the city.

\section{Methodology}
\label{methodology}

In this section, we introduce the details of our proposed framework for the urban graph representation learning task. First, we show the overview of our framework and outline each module. Then, we discuss the CGAP mechanism which aims at integrating the information from the local regions and the global graph.  Finally, we present the learning objective function and its related urban computing downstream tasks. 

\subsection{Framework Overview}

\begin{figure}[htbp]
\vspace{-5pt}
    \centering
    \includegraphics[width=0.45\textwidth]{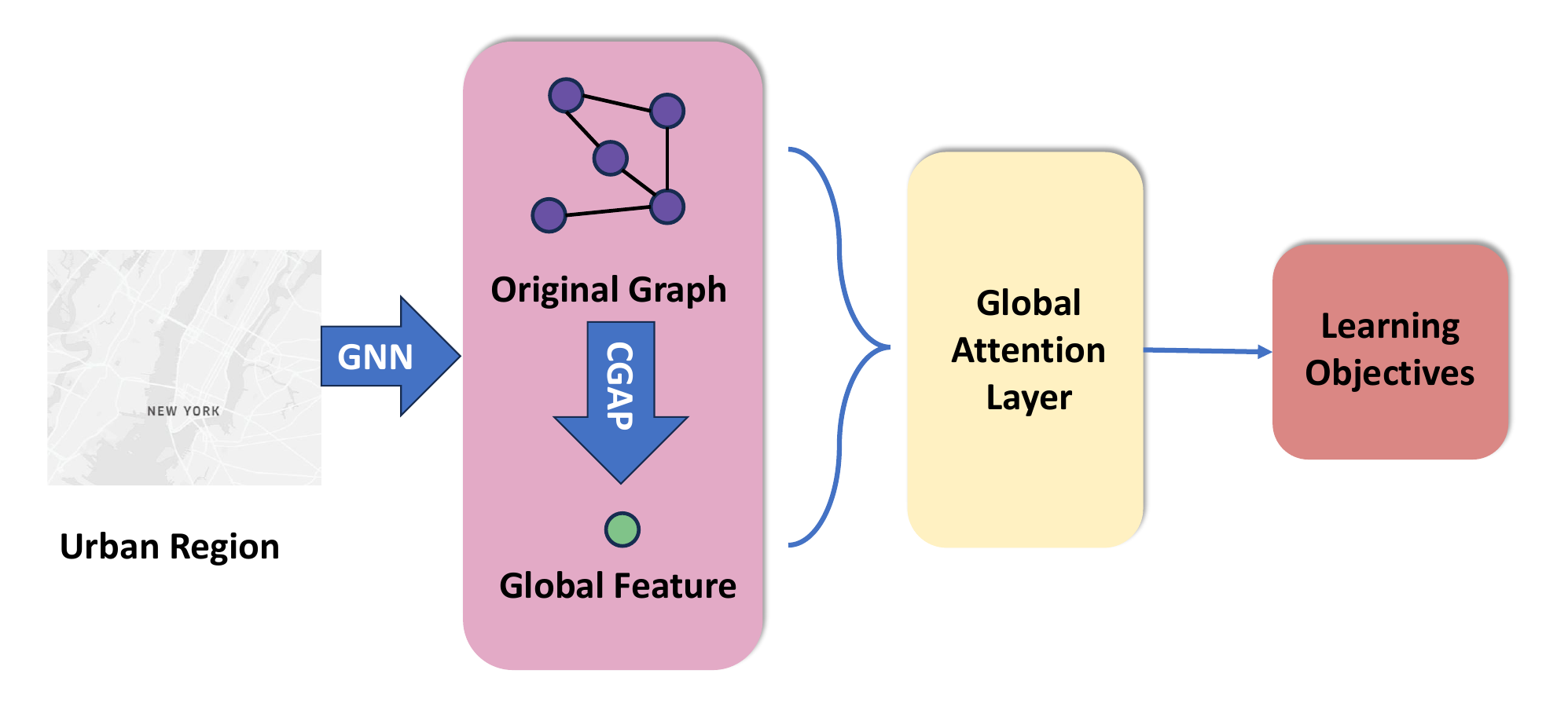}
    \caption{Our framework for urban graph representation learning. Through GNN, the original  urban graph embedding is initialised. And then model gets the global feature through the CGAP mechanism. Global attention layer integrates the original graph embedding and global feature to strengthen region representation. Finally, we set multiple learning objectives to train the model. In multiple objectives, we focus not only  region embedding effectiveness but model's performance in downstream tasks.}
    \label{fig:framework}
    \vspace{-5pt}
\end{figure}

Our model is illustrated in Figure~\ref{fig:framework}.  After initializing urban region information into an original graph, GNNs with CGAP mechanism utilize the original graph to extract various global features. In order to get the local and global features in a graph, the global attention layer integrates compressed feature nodes and the whole graph into an urban region graph embedding result. Based on the urban computing-related objectives, we can train our model to learn effective urban region representations. 

\subsection{CGAP mechanism}

\begin{figure}[htbp]
    \centering
    \includegraphics[width=0.45\textwidth]{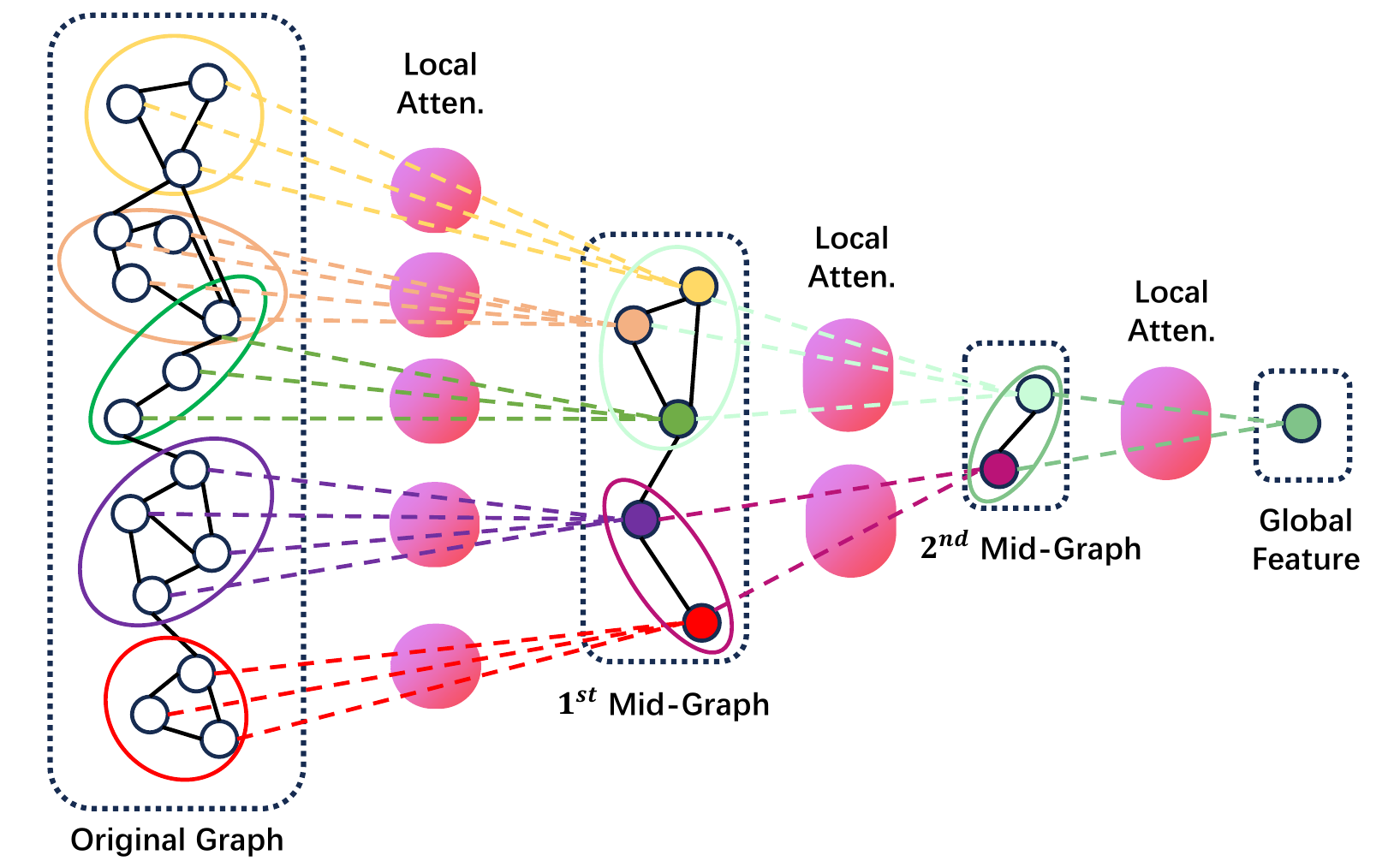}
    \caption{The illustration of CGAP module in our framework. This pooling process aims at integrating local nodes and generating global feature. Original graph from GNN embedding initialisation is aggregated into a global feature node through multi-layer local attention pooling process. Each attention unit has its own corresponding local region nodes which are aggregated into a coarsened node. }
    \label{fig:CGAP}
    \vspace{-5pt}
\end{figure} 

To aggregate the original graph into a coarsened graph and thus extract the feature nodes,  we propose the GNNs with Coarsened Graph Attention Pooling (\textbf{CGAP} mechanism) shown in Figure ~\ref{fig:CGAP}. Its proposal aims at making up for GNN architecture's deficiencies in dealing with urban reign. With the coarsened graph and global feature node generation process, the relation among regions is deepening and local regions in urban are becoming a node that represents interaction information. Therefore, the extracted global feature node which has global information could be provided to enhance region embedding. We first describe GNNs in region graph embedding initialization. And then we present the graph aggregation process with the local attention units.

\subsubsection{Graph Region Embedding Initialisation} 
In this work, we convert $\mathcal{G} = (\mathcal{V},\mathcal{E})$ into $\mathcal{G} = (A, F)$, where $A \in \{0, 1\}^{\|\mathcal{V}\| \times \|\mathcal{V}\|}$  is the adjacency matrix, and $F \in \mathbb{R}^{\|\mathcal{V}\| \times d}$ is the node feature matrix where each node has $d$ features. To capture useful region feature and relation, we build upon GNNs~\cite{GCN} which apply the hierarchical structure to pass message. Given the random initial node feature embedding $F$, we update the feature embedding through the following message-passing architecture.
\begin{equation}
    \begin{split}
    H^{(k)} &= \mathrm{M}(A, H^{(k-1)}, W) \\
    &= \mathrm{ReLU}(\tilde{D}^{- \frac{1}{2}}\tilde{A}\tilde{D}^{- \frac{1}{2}}H^{(k-1)}W^{(k-1)}), \\
    \end{split}
\end{equation}
where $H^{(k)} \in \mathbb{R}^{\|\mathcal{V}\| \times d}$ is the hidden node embedding from the $k$ step in GNN, $\mathrm{M}$ is the message propagation function, and $W \in \mathbb{R}^{d \times d}$ is the trainable matrix. In $\mathrm{M}$ function, $\tilde{A} = A + I, \tilde{D} = \sum_{j}\tilde{A}_{ij}, H^{(0)} = F$, and $W^{(k-1)}$ is the $k$ step training weight matrix.    

For simple description in pooling process, we replace the GNNs calculation with $Z = \mathrm{GNN}(A,F)$ which represents the operation of multiple message passing and iteration  through GNNs. 

\subsubsection{Pooling Process} The pooling process is shown in Figure ~\ref{fig:CGAP}. Neighbor nodes are aggregated into a coarsened node to represent local information containing these regions. Through the multi-layer local attention modules, we can get the coarsened graphs. Finally, we can get the urban global feature node as CGAP's output. Given the initial embedding result $Z$, this hierarchical pooling iteration process can be described as follows:
\begin{equation}
    \begin{aligned}
    &h_0 = Z, \\
    &h_{l+1} = \mathrm{Update}(h_l; \theta_{l}), 
    \end{aligned}
\end{equation}
where $h_l$ represents the middle graph embedding and $\theta_{l}$ is the parameters of layer $l$. Finally, we can get the global feature node embedding$h_g$.

\subsubsection{Local Attention Unit} We define the assignment attention weight matrix $S^{l} \in \mathbb{R}^{n_{l} \times n_{l+1}}$ where $n_l$ is the account of mid-graph nodes at layer $l$. $S^{l}$ provides a method to assign local region nodes into a coarsened node in the next layer. Given the region embedding $Z^{l}$ from layer $l$, we apply the following function to get middle matrix $X \in \mathbb{R}^{n_{l+1} \times d}$:
\begin{equation}
    X = \sum_{i=1}^{\|\mathcal{A}_l\|} \alpha^{l}S^{l^{T}}_{i}Z^{l},
\end{equation}
where $\|\mathcal{A}_l\|$ is the account of local attention at layer $l$,  $\alpha^{l}$ is the hyperparameter of layer $l$ and $Z^{0}$ is the initialisation result. In the local attention unit, we mask the nodes which are not in the local regions to ensure its focus scope. In the pooling process, each unit has its corresponding local regions.

To ensure the next layer node embedding learning the local relation, we also map the adjacency matrix information and calculate it in the unit: 
\begin{equation}
    A^{l+1} = \sum_{i=1}^{\|\mathcal{A}_l\|} \alpha^{l}S^{l^{T}}_{i}A^{l}S^{l}_{i},
\end{equation}
where $A^{l+1} \in \mathbb{R}^{n_{l+1} \times n_{l+1}}$, and $A^0$ is the original adjacency matrix. 

When the local attention unit updates $Z^l$ to $Z^{l+1}$, the interaction of local graph is based on $X$ and its adjacency matrix $A^{l+1}$. We consider the node embedding information mapped in the same latent space, and thus transform coarsened nodes by the following equation:
\begin{equation}
    Z^{l+1} = \mathrm{ReLU}(A^{l+1}X \textbf{W}^l + \textbf{b}^{l}),
\end{equation}
where $\textbf{W}^l, \textbf{b}^{l}$ are the layer parameters which allow the result from local attention units to map in the middle graph latent space. We obtain the local graph information and build the middle graph composed of nodes to avoid flat problem due to increasing number of GNN layers. Through multi-layer pooling, we can get the single node embedding $h_g$ as the global feature.

\subsubsection{Global Attention Layer} To capture both regional and global features from the CGAP mechanism,  we apply a global attention method that integrates information to compute the region representations. Based on the global feature $h_g$, our framework focuses on the important region feature which has a strong association with urban.  

Formally, given the result $h_g$ from CGAP module and the region initial embedding $Z$, we compute the global attention to enhance region representation $\hat{E}$ as follows.

\begin{equation}
\begin{aligned}
    &K_w = (\|_{i=1}^{\|\mathcal{V}\|}h_g) \mathbf{W_k},\\
    &Q_w = Z\mathbf{W_q},  V_w = Z\mathbf{W_v},\\
    &\mathrm{Atten}(K, Q, V) = \mathrm{softmax}(\frac{QK^T}{\sqrt{d}})V,\\
    &\hat{E} = \mathrm{Atten}(K_w, Q_w, V_w),
\end{aligned}
\end{equation}
where $\mathbf{W_k}, \mathbf{W_q}, \mathbf{W_v}$ are learned parameters, and $\|$ is the operation to concatenate vectors into a matrix.  

\subsection{Learning Objectives }
In training our model, we integrate region embedding effectiveness improvement and multi-task learning to formulate our objective function. We acknowledge the importance of enhancing global feature addition for the learning process. Additionally, by leveraging region representation for downstream tasks, we enable the downstream effectiveness to align with the learning function. Subsequently, we introduce distinct region embedding loss and multi-task loss functions.

\subsubsection{Region Embedding Loss} To measure the influence of global feature addition, we define the region embedding loss function $\mathcal{L}_{r}$. Given the original graph embedding $Z$ and the final result $\hat{E}$, $\mathcal{L}_{r}$ can be computed  by the following equation:
\begin{equation}
    \mathcal{L}_{r} =  \sum_{i=1}^{\|\mathcal{V}\|}  \exp(-\|\hat{e}_i - z_i \|_2 ),
\end{equation}
where $\hat{e}_i, z_i$ is the $i$ region vector from  $\hat{E}, Z$, and $\|\cdot\|_2$ is the euclidean distance function. 

\subsubsection{Multi-Task Loss} We utilize various urban data to train our model, selecting mobility data and Points of Interest (POIs) based on region relations and attributes to calculate the loss function. The region representation is then decoded for downstream tasks based on their respective loss functions.

For tasks involving mobility and POI prediction, we employ Linear modules to decode the region representations. Given the region representation $\hat{E}$, these modules generate the predicted mobility $\hat{M}$ and region POI embedding $\hat{P}$ as follows.

\begin{equation}
    \begin{aligned}
        &\hat{M} = \mathrm{Linear}_{m}(\hat{E}),  \hat{P} = \mathrm{Linear}_{p}(\hat{E}),\\
    \end{aligned}
\end{equation}
where $\hat{M} \in \mathbb{R}^{\|\mathcal{V}\| \times \|\mathcal{V}\|}, \hat{P} \in \mathbb{R}^{\|\mathcal{V}\| \times d}$. 

Given the human mobility adjacency matrix $M = \{m_{ij}\}$ and its prediction $\hat{M}$, we can compute the probability in mobility distribution as follows.

\begin{equation}
\begin{aligned}
    &Pr(j|i) = \frac{m_{ij}}{\sum_{k=i}^{\|\mathcal{V}\|} m_{ik}},\\
    &\hat{Pr}(j|i) = \frac{\hat{m}_{ij}}{\sum_{k=i}^{\|\mathcal{V}\|} \hat{m}_{ik}},\\
\end{aligned}
\end{equation}
where $Pr(j|i)$ is the probability where resident source is region $i$ and destination is region $j$, $\hat{Pr}(j|i)$ is the model prediction result.

And then we make use of cross entropy loss function to evaluate the proximity of real mobility to the predicted outputs: 
\begin{equation}
    \mathcal{L}_{mob} = \sum_{i=1}^{\|\mathcal{V}\|} \sum_{j=1}^{\|\mathcal{V}\|} - Pr(j|i)\log(\hat{Pr}(j|i)).
\end{equation}

To retain the relation among urban regions, we adopt the similarity of region attribute, and design POI loss function based on its embedding result. Given the region attributes $P = \{p_1, p_2, ..., p_{\|\mathcal{V}\|}\}$ and POI embedding $\hat{P} =  \{\hat{p}_1, \hat{p}_2, ..., \hat{p}_{\|\mathcal{V}\|}\}$, we formalize the POI loss $\mathcal{L}_{poi}$ as follows.

\begin{equation}
    \mathcal{L}_{poi} = \sum_{i=1}^{\|\mathcal{V}\|} \sum_{j=1}^{\|\mathcal{V}\|} (\frac{p_i}{p_j} - \hat{p}_{i}^{T}\hat{p}_{j})^{2}.
\end{equation}

As the total result, the loss value $\mathcal{L}$ can be computed as follows.  

\begin{equation}
    \mathcal{L} = \beta \mathcal{L}_{r} + (1 - \beta)(\mathcal{L}_{mob} + \mathcal{L}_{poi}),
\end{equation}
where $\beta$  is hyper parameter to assign the loss weights in $\mathcal{L}$. 

\subsubsection{Discussion}
\textbf{Global Urban Pooling and Ours.} Compared to the traditional frameworks for learning region embeddings, our method adopts local node clustering and local attention units to learn the sampling so that the model can decide which information is worth sampling. Traditional methods and GNNs adopt the global urban graph in the pooling process. We summarize the advantages of our structure which global methods do not have as follows.

\begin{itemize}
    \item Constrained by regional nodes, our local attention units focus on information exchange within the immediate neighborhood, avoiding the over-smoothing issue associated with high-order neighborhood information in Graph Neural Networks (GNNs).

    \item Each unit's receptive field is confined within a cluster, allowing for the preservation of heterogeneous information pertaining to the topology and features of urban regions.
\end{itemize}

\noindent \textbf{Analyses.} We also analyze the theoretical complexity of the pooling process. Given the input graph with node set $V$ and edge set $E$ in the pooling layer, our proposed CGAP requires storage complexity of $\mathcal{O}(\frac{1}{\mu}\|V\|^2)$ where one cluster has $\mu$ nodes, as region matrix is sparse. DiffPooling requires  $\mathcal{O}(\|V\|^2)$. And SAGPool requires $\mathcal{O}(\|V\|+\|E\|)$, where $\|E\| \approx \|V\|^2$ in urban computing because of the high complexity of the urban structure. In the time complexity calculation, we set the pooling depth to $k$. In this condition, CGAP requires $\mathcal{O}(k\mu\log\frac{\|V\|^2}{\mu})$ whereas DiffPool requires $\mathcal{O}(k\log\|V\|^2)$. So the complexity of our model is acceptable.

\section{Experiments}
In this section, we describe the details of our experiments. First, we introduce the dataset and implementation information related to the experiment settings. Then, we present the baseline models and downstream tasks in experiments. Under this condition, we evaluate the performance of our model and baseline models. At last, we analyse the results of our model and current methods, and validate the effectiveness of our approach. 

\subsection{Datasets}
We collect the real-world urban data from New York City on NYC open data \footnote{https://opendata.cityofnewyork.us}datasets. Specially, we focus urban regions in  Manhattan, and apply taxi trips as resident mobility. The regional division is base on the community boards. The detailed description of datasets id shown in Table ~\ref{tab:dataset_description}.

\begin{table}
\centering
\resizebox{.95\columnwidth}{!}{
\begin{tblr}{
  cells = {c},
  hline{1-2,7} = {-}{},
}
\textbf{Dataset} & \textbf{Description}                                    \\
Regions          & 180 regions in~Manhattan, NYC~                          \\
POI data         & Around 20 thousand POIs including station, stores, etc. \\
Taxi trips       & Around 10 million taxi trip records during one month    \\
Check-in data    & Over 100 thousand check-in locations                    \\
Crime data       & Around 40 thousand crime records~during one year        
\end{tblr}}
\caption{Dataset Description}
\label{tab:dataset_description}
\end{table}

\subsection{Experiment Settings}
In order to compare methods fairly, we use the Adam optimizer with the same learning rate of $1e-3$ in our experiments. And neural network models are trained in 2000 epochs. During the training, we set the hidden dimension of models to 128, and the dropout to 0.5. The experiments were performed on the GeForce RTX 2080 Ti with 11G memory. 

To align AI methods with the UN Sustainable Development Goals, we select crime prediction, check-in prediction, and land usage classification as downstream tasks. The crime prediction task contributes to Peace, Justice, and Strong Institutions, while the check-in prediction and land usage classification tasks are associated with Sustainable Cities and Communities.
\subsection{Baseline Models}
We compare our model with the following baseline approaches.
\begin{itemize}
    \item \textbf{GAE}~\cite{GAE} uses a GCN encoder and an inner product decoder to learn interpretable latent representations for undirected graphs.
    
    \item \textbf{Node2Vec}~\cite{node2vec} maps the graph nodes into a low-dimensional space to maximize the likelihood of preserving network neighborhoods of nodes, and propose a biased random walk process which explores diverse neighborhoods.

    \item \textbf{HDGE}~\cite{HDGE} jointly learns the representations from a traffic flow graph and a spatial graph, and uses it to  measure the relationship strengths between regions.

    \item \textbf{ZE-Mob}~\cite{ZEMob} learns region embeddings from the co-occurrence in human mobility data. And the model incorporate multiple mobility data into the modeling of zone embeddings.

    \item \textbf{MV-PN}~\cite{MVPN} constructs multi-view POI-POI networks to represent regions, and introduces spatial autocorrelations and top-k locality into region embedding.

    \item \textbf{MVURE}~\cite{MVURE} adopts the intra-region and inter-region data to construct multi-view graphs, and applies joint learning module to learn region embedding.

    \item \textbf{MGFN}~\cite{MGFN} focuses mobility patterns by human mobility, and adopts a mobility graph fusion module and the mobility pattern joint learning module to learn the embedding.

    \item \textbf{HREP}~\cite{HREP} proposes heterogeneous region embedding (HRE) with relation-aware GCN and prompt learning for downstream tasks to address both intra-region and inter-region correlations.

    \item \textbf{DiffPooling}~\cite{flat} uses our framework which replace global feature extraction with DiffPooling method.

    \item \textbf{SAGPool}~\cite{selfattention} uses self-attention, considering both node features and graph topology. 
    
    \item \textbf{ASAP}~\cite{asap}learns a sparse soft cluster assignment for nodes at each layer.

    \item \textbf{Ours(CGAP)} uses our framework with CGAP mechanism.
    
\end{itemize}

\subsection{Experimental Results}
\begin{figure}
    \centering
    \subfigure[NMI]{\includegraphics[width=.48\columnwidth,height=0.13\textwidth]{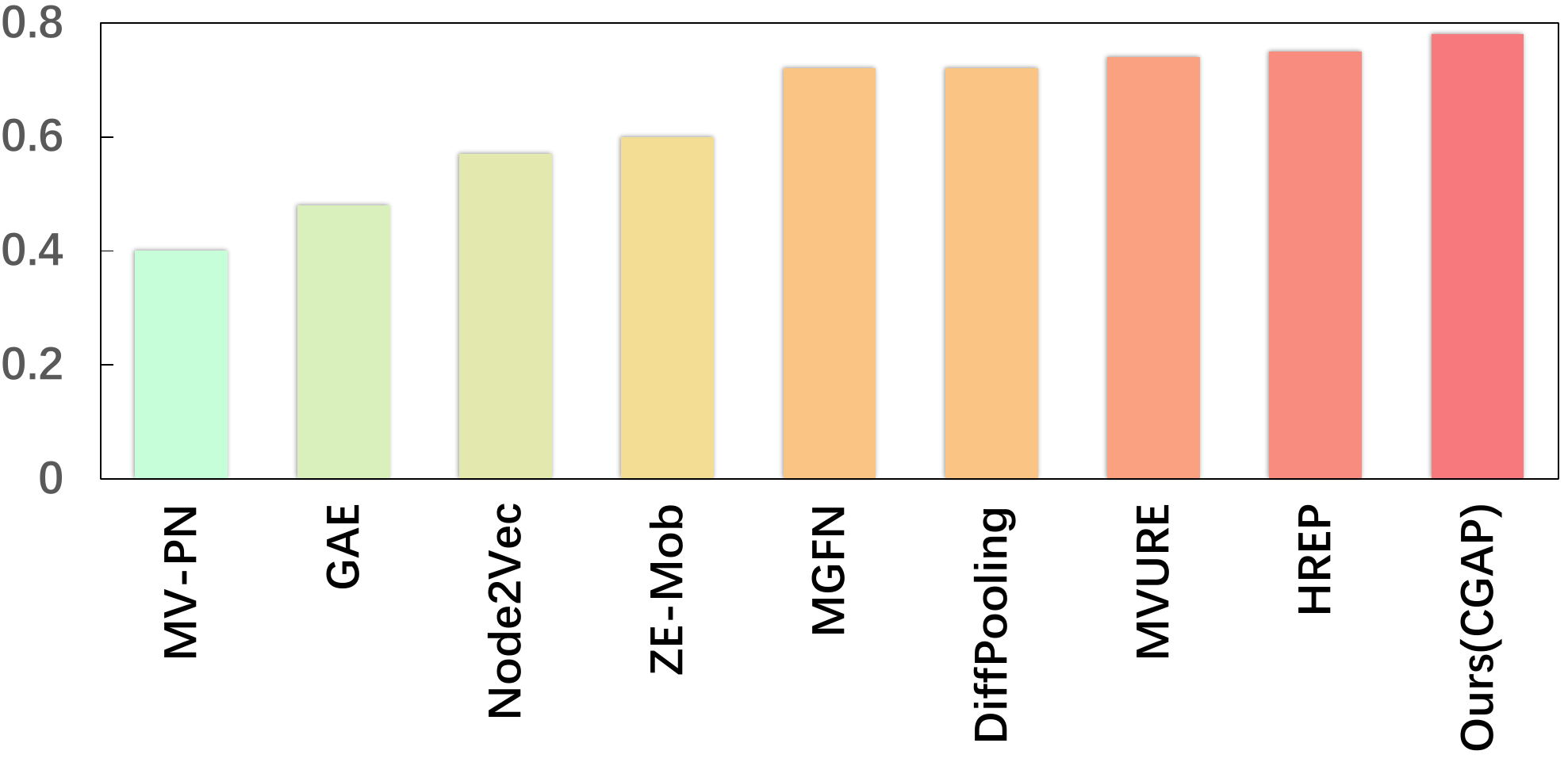}}\subfigure[ARI]{\includegraphics[width=.48\columnwidth,height=0.13\textwidth]{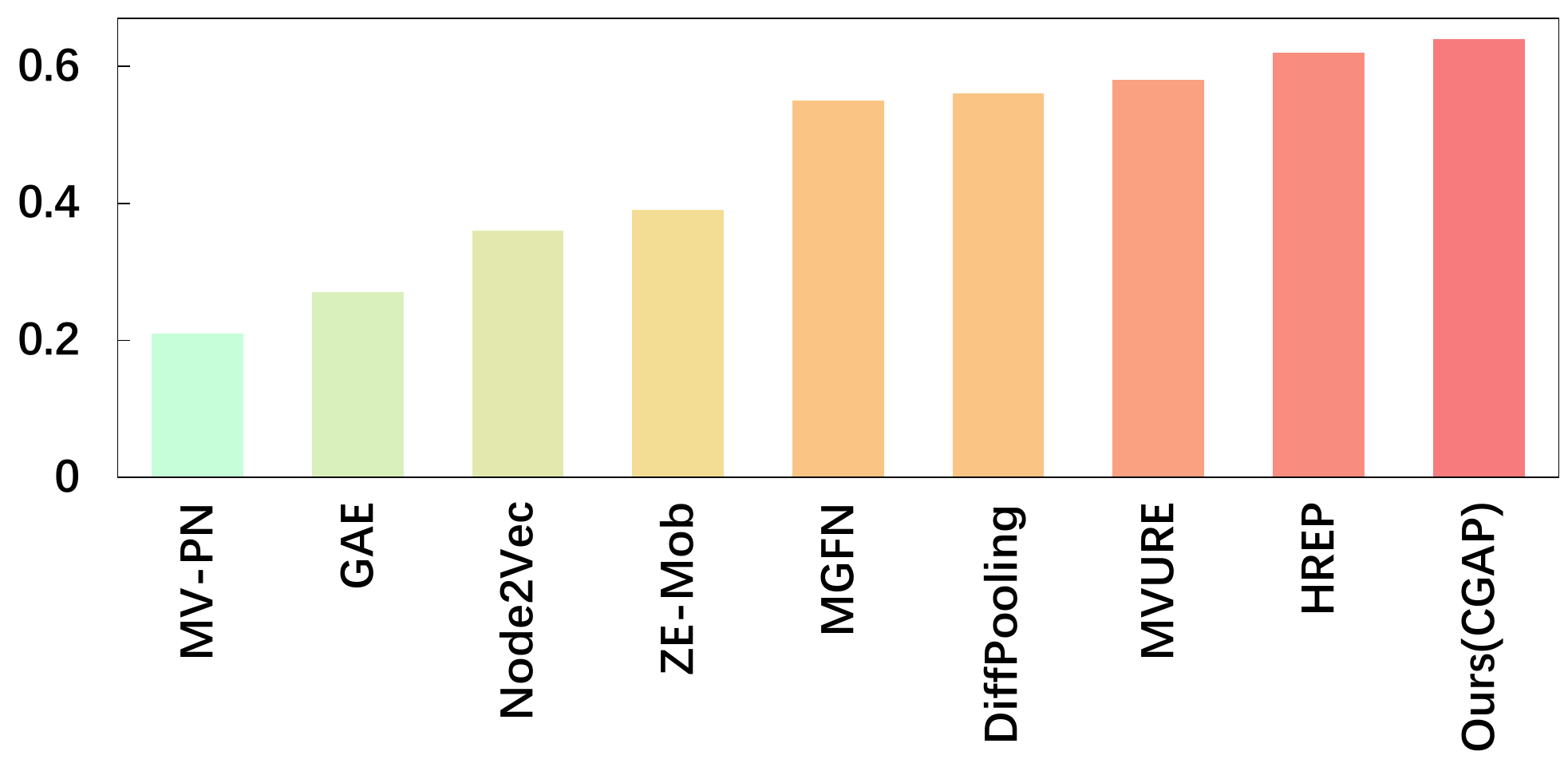}}
    
\caption{Land usage classification results}
\label{fig:landuse}
\end{figure}

In our experiments, we selected two downstream tasks, crime prediction and check-in prediction, for the main comparison experiments. We used several standard metrics to evaluate performance: Root Mean Squared Error (RMSE), Mean Absolute Error (MAE), and $R^2$, which are commonly used in regression tasks. We also applied the Lasso regression model ~\cite{lasso} for prediction. The performance of the comparison experiment is shown in Table ~\ref{tab:result}. Our framework with the CGAP mechanism outperforms all state-of-the-art methods. Additionally, we evaluated method performance in a land usage classification task, using two metrics: Normalized Mutual Information (NMI) and Adjusted Rand Index (ARI). The results, shown in Figure ~\ref{fig:landuse}, indicate that our framework with the CGAP mechanism also performs well in these experiments.

From the experiment results, we deduced that GNN architectures used in urban region embedding have limitations in neighbor information extraction. Encoders like Node2Vec, which perceive neighbor nodes, perform better than methods like GAE, which focuses only on the embedding nodes. Fusion methods like MVURE and HREP, which integrate local urban information such as inter-region data, improve task performance. Compared to DiffPooling where pooling is done using an aggregation method for the graph as a whole, our approach focuses on the local graph information where local attention units are created for the specific local regions. Generally, our approach balances the hierarchical pooling approach with the task characteristics of region embedding.
Among all approaches, our method performs best. We not only enhance the GNN architecture with local attention units that fuse local urban information but also add a global feature to region embedding. To validate these model modules, we designed ablation experiments. 

From the perspective of data usage, methods focusing on mobility data or POI data, such as ZE-mob, MV-PN, and MGFN, face limitations in performance improvement. Specifically, methods using mobility, like MGFN, achieve better results among approaches with a single data type, even surpassing the multi-view data approach, MVURE. Therefore, we conducted ablation experiments on data to validate the effectiveness of multi-view data in the ablation study.

\begin{table}
\centering
\resizebox{.95\columnwidth}{!}{
\begin{tblr}{
  cells = {c},
  cell{1}{2} = {c=3}{},
  cell{1}{5} = {c=3}{},
  hline{1,3,14} = {-}{},
  hline{2} = {2-7}{},
}
                    & Crime Prediction &                &               & Check-in Prediction &                 &               \\
                    & MAE              & RMSE           & $R^{2}$        & MAE                 & RMSE            & $R^{2}$        \\
GAE                 & 96.55            & 133.10         & 0.19          & 498.23              & 803.34          & 0.09          \\
Node2Vec            & 75.09            & 104.97         & 0.49          & 372.83              & 609.47          & 0.44          \\
ZE-Mob              & 101.98           & 132.16         & 0.20          & 360.71              & 592.92          & 0.47          \\
MV-PN               & 92.30            & 123.96         & 0.30          & 476.14              & 784.25          & 0.08          \\
MVURE               & 76.43            & 99.03          & 0.55          & 343.53              & 538.15          & 0.57          \\
MGFN                & 72.61            & 93.43          & 0.60          & 328.22              & 494.63          & 0.63          \\
HREP                & 71.02            & 90.91          & 0.62          & 310.86              & 493.86          & 0.63          \\
DiffPooling   & 73.17            & 93.75          & 0.59          & 323.98              & 515.24          & 0.60          \\
SAGPool             & 73.24            & 94.24          & 0.58          & 316.33              & 498.57          & 0.60          \\
ASAP                & 70.52            & 90.52          & 0.63          & 305.49              & 486.43          & 0.62          \\
\textbf{Ours(CGAP)} & \textbf{68.10}   & \textbf{87.03} & \textbf{0.65} & \textbf{299.81}     & \textbf{471.96} & \textbf{0.66} 
\end{tblr}}
\caption{Comparison on crime prediction and check-in prediction task.}
\label{tab:result}
\end{table}

\subsection{Sensitivity Analysis}
\begin{figure}[htbp]
\vspace{-5pt}
    \centering
    \includegraphics[width=.7\columnwidth]{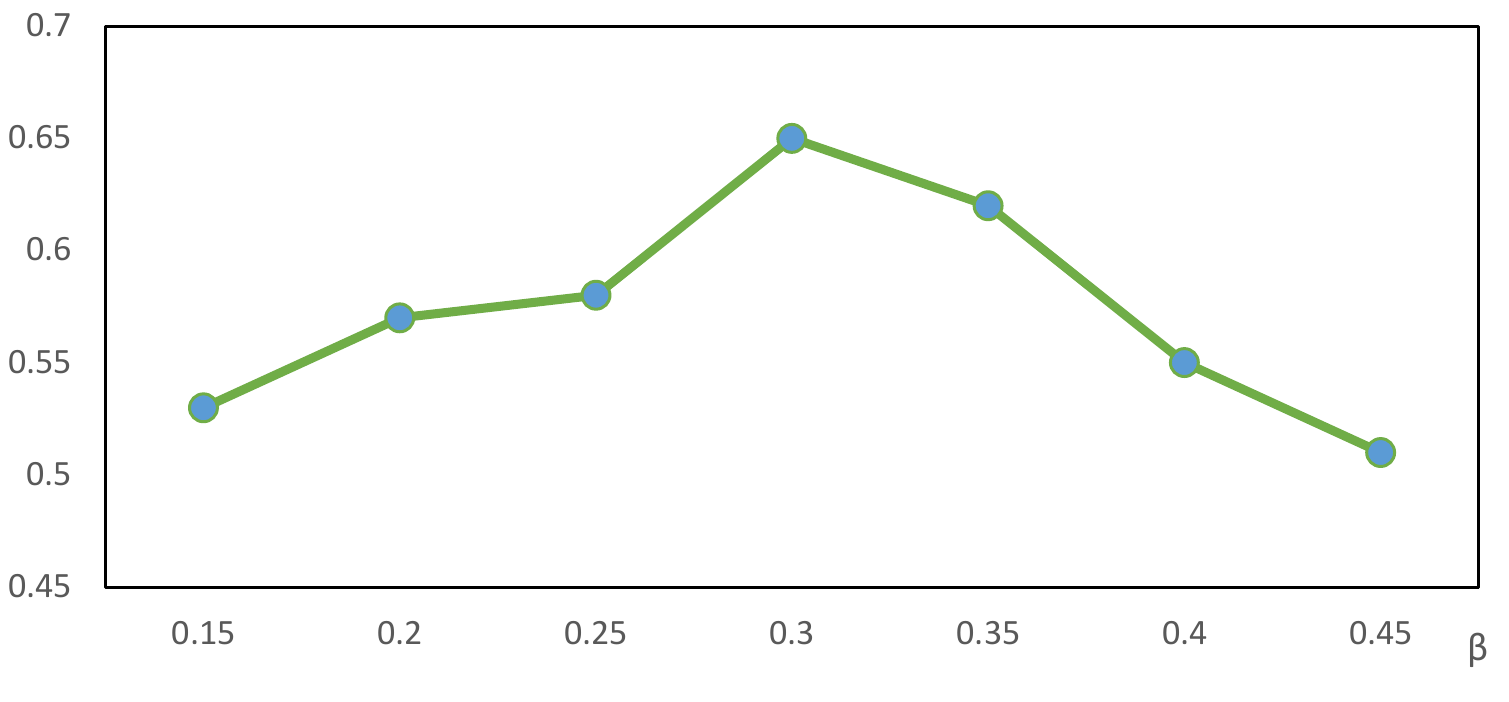}
    \caption{Sensitivity Analysis in different hyperparameter values, where we observe the effect of different $\beta$ on $R^2$ under the crime prediction task. }
    \label{Sensitive}
    \vspace{-5pt}
\end{figure}

We analyze the sensitivity of our model in the task of crime prediction. In this experiment, we utilize $R^2$ as evaluation metric. The result is shown in Figure ~\ref{Sensitive}. We set hyperparameter $\beta$ from 0.15 to 0.45, and find when $\beta$ is 0.3, the performance of our framework is better than other circumstances.  

\subsection{Ablation Study}
\label{sec:ablation}
To verify the effectiveness of our modules in the framework, we conduct ablation experiments about the CGAP mechanism. We focus on the effectiveness of local attention units and global features in our framework. Local attention units afford fusion functions to integrate local information, and global feature provides generalized information. Therefore, we discuss two circumstances of the framework: 1) \textbf{CGAP(L)} replacing local attention units with Linear modules; 2) \textbf{CGAP(No-G)} without global feature input for global attention layer. The result of ablation experiments related to the CGAP module is shown in Table ~\ref{tab:abacgap}.    

\begin{table}
\centering
\resizebox{.95\columnwidth}{!}{
\begin{tblr}{
  row{even} = {c},
  row{3} = {c},
  row{5} = {c},
  cell{1}{2} = {c=3}{c},
  cell{1}{5} = {c=3}{c},
  hline{1,3,6} = {-}{},
  hline{2} = {2-7}{},
}
                    & Crime Prediction &                &               & Check-in Prediction &                 &               \\
                    & MAE              & RMSE           & $R^{2}$        & MAE                 & RMSE            & $R^{2}$        \\
CGAP(L)             & 80.61            & 103.43         & 0.51          & 335.96              & 528.94          & 0.58          \\
CGAP(No-G)          & 76.43            & 99.03          & 0.55          & 343.53              & 538.15          & 0.57          \\
\textbf{Ours(CGAP)} & \textbf{68.10}    & \textbf{87.03} & \textbf{0.65} & \textbf{299.81}     & \textbf{471.96} & \textbf{0.66} 
\end{tblr}}
\caption{Ablation experiments related to CGAP.}
\label{tab:abacgap}
\end{table}

From this table, we can observe that local attention units effectively provide a fusion method to improve model performance. Through multi-layer compression, the relation among regions is strengthened. Global feature node is introduced to model and makes region representation learn from compression graph.  

In order to clarify the multi-view data function, we also conduct the ablation experiments about data used in the main comparison experiments. We divide the training datasets into two circumstances: 1) \textbf{Ours(POI)} including single POI data; 2) \textbf{Ours(mobility)} including single mobility data. The result of the ablation experiments about multi-view data is shown in Table ~\ref{tab:abamul}.

With the addition of mobility data, the experimental performance has a huge boost. The addition of POI data is useful but limited. Combining these two region data is the best choice to train region representation learning for downstream tasks.

\begin{table}
\centering
\resizebox{.95\columnwidth}{!}{
\begin{tblr}{
  cell{1}{2} = {c=3}{},
  cell{1}{5} = {c=3}{},
  hline{1,3,6} = {-}{},
  hline{2} = {2-7}{},
}
                    & Crime Prediction &                &               & Check-in Prediction &                 &               \\
                    & MAE              & RMSE           & $R^{2}$        & MAE                 & RMSE            & $R^{2}$       \\
Ours(POI)           & 108.79           & 140.08         & 0.11          & 451.64              & 692.89          & 0.28          \\
Ours(mobility)      & 73.85            & 94.97          & 0.58          & 318.08              & 509.37          & 0.61          \\
\textbf{Ours(CGAP)} & \textbf{68.10}    & \textbf{87.03} & \textbf{0.65} & \textbf{299.81}     & \textbf{471.96} & \textbf{0.66} 
\end{tblr}}
\caption{Ablation experiments related to multi-view data.}
\label{tab:abamul}
\end{table}

\section{Related Work}
\subsection{Urban Region Embedding}
The proliferation of mobile devices and rapid urbanization improve the development of cities. A large number of researchers~\cite {luo2022urban,HREP,MVURE} pay attention to urban computing tasks and analyze the city with a machine. Since cities afford the daily lives of their inhabitants, urban computing tasks such as traffic control~\cite{traffic} are closely related to the daily functions of cities. 

With the functional subdivision of different areas in the city, urban region representation learning rises from various urban computing tasks. Social characteristics of the region such as POI and resident mobility affect region embedding. As a typical indicator of regional prosperity, POI is an indispensable factor for representation learning.  Researches ~\cite{MVPN}  propose a POI-based embedding strategy and network to leverage region properties. Besides POI, other region attributes such as street information~\cite{luo2022geo} are used in this task. Compared with POI, mobility data is widely used in the embedding task related to urban spatial graphs. As the mobility data has spatio-temporal items, this feature is assisted in representing regions. \citeauthor{ZEMob} \shortcite{ZEMob} pursuits to find co-occurrence in human mobility and add it into the embedding process. \citeauthor{MGFN} \shortcite{MGFN} introduces the graphs with spatio-temporal similarity as mobility patterns for joint learning.

\subsection{Graph Representation Learning}
Graph embedding aims to transform node attributes into a lower-dimensional space, generating vector representations that effectively capture node relationships~\cite{cui2018survey}. Recent research has centered on Graph Neural Networks (GNNs), extensively utilized for structured graph data~\cite{wu2020comprehensive}. Notably, Graph Convolutional Networks (GCNs)~\cite{GCN} integrate convolutional techniques from computer vision into GNNs, significantly influencing learned representations. Various strategies have emerged to enhance GCN performance. For instance, in spatial-temporal prediction tasks, STGCN~\cite{temporalGNN_0} leverages GCNs' adaptable propagation mechanism to learn node features. Furthermore, the Graph Attention Network (GAT)~\cite{velickovic2017graph} refines neighbor aggregation through attention mechanisms.

Graph pooling is a fundamental element in GNN architectures~\cite{poolingre}. It reduces dimensionality and compresses the input feature map for computational efficiency. Basic pooling methods~\cite{xu2018powerful} aggregate node representations through flattening techniques like summing or averaging node embeddings. Advanced pooling techniques~\cite{flat,diffpooling_atten} refine graph representations across multiple network layers. Notably, \citeauthor{diffpooling_atten} \shortcite{diffpooling_atten} introduces local pooling and node attention mechanisms in each layer. 

\section{Conclusion}
In this paper, we introduce a novel mechanism, \textbf{CGAP}, designed for node aggregation and global feature extraction within urban region graphs. It efficiently condenses specified local region graphs into singular nodes using a local attention unit, addressing the challenge of node aggregation in urban contexts. Furthermore, to overcome the inherent flatness problem in graph neural networks, CGAP employs a hierarchical structure. This structure not only preserves the original graph information but also integrates a global feature node into the global attention layer, enhancing the model's ability to capture comprehensive urban dynamics. Specifically, CGAP leverages data on human mobility and POIs to construct detailed region attributes and relationships within the graph architecture, facilitating a deeper understanding of region characteristics. Our experiments, utilizing real-world datasets for downstream applications, demonstrate that CGAP significantly surpasses all baseline methods in performance. Aligned with the UN SDGs, we aim to broaden our framework to include more downstream tasks, delving deeper into the capabilities of CGAP to enhance urban data analyses.

\section*{Acknowledgements}
This work was supported by the National Natural Science Foundation of China under Grant No. 62106274 and the Fundamental Research Funds for the Central Universities, Renmin University of China under Grant No. 22XNKJ24. We also wish to acknowledge the support provided by the Intelligent Social Governance Platform, Major Innovation \& Planning Interdisciplinary Platform for the "Double-First Class" Initiative. Xiao Zhou is the corresponding author.

\bibliographystyle{named}
\bibliography{ijcai24}

\end{document}